\documentclass[12pt,peerreview]{IEEEtran}

\usepackage{graphicx}
\usepackage{cite}
\usepackage{amsmath}
\usepackage{amssymb}
\usepackage{array}
\usepackage{dcolumn}
\usepackage{mathrsfs}
\usepackage{caption}

\newtheorem{theorem}{Theorem}

\newtheorem{definition}{Definition}

\newcommand{\dd}{{\rm d}}
\newtheorem{remark}{Remark}

\newcommand{\sd}{Schr\"{o}dinger }

\newcommand{\II}{\mathbf{I}}
\newcommand{\JJ}{\mathbf{J}}
\newcommand{\bb}{\hat{\mathbf{b}}}
\newcommand{\cc}{\hat{\mathbf{c}}}
\newcommand{\ee}{\mathbf{e}}
\newcommand{\ii}{\mathbf{i}}
\newcommand{\jj}{\mathbf{j}}

\newcommand{\kk}{\mathbf{k}}
\newcommand{\pp}{{\mathbf{p}}}
\newcommand{\PP}{\mathbf{P}}
\newcommand{\QQ}{\mathbf{Q}}
\newcommand{\GG}{\mathbf{G}}
\newcommand{\DD}{\mathbf{D}}
\newcommand{\CC}{\mathbf{C}}

\newcommand{\EE}{\mathbf{E}}
\newcommand{\FF}{\mathbf{F}}

\newcommand{\LL}{\mathbf{L}}
\newcommand{\MM}{\mathbf{M}}
\newcommand{\NN}{\mathbf{N}}
\newcommand{\HH}{{\mathbb{D}}}
\newcommand{\ww}{\hat{\mathbf{w}}}
\newcommand{\xx}{\hat{\mathbf{x}}}

\newcommand{\aaa}{\hat{\mathbf{a}}}

\newcommand{\tr}{{\rm Tr}}%
\newcommand{\w}{{\omega}}%

\begin{document}
\title{Signal Flows in Non-Markovian Linear Quantum Feedback Networks}
\author{Re-Bing Wu ~\IEEEmembership{Member,~IEEE}
, Jing Zhang ~\IEEEmembership{Member,~IEEE}, Yu-xi Liu and Tzyh-Jong Tarn ~\IEEEmembership{Life~Fellow,~IEEE},
\thanks{This research was supported in part by the National Natural Science
Foundation of China under Grant Number 61374091 and 61134008.}
\thanks{RBW and JZ are with Department of Automation, Tsinghua University,
Beijing, 100084, China; Yu-xi Liu is with Institute of Microelectronics, Tsinghua University,
Beijing, 100084, China; and TJT is with Department of Electrical and Systems Engineering,
Washington University in St. Louis, St. Louis, MO 63130, USA. All authors are with the Center for Quantum Information Science and Technology, TNList, Beijing, 100084, China; }
}
\maketitle

\begin{abstract}
Enabled by rapidly developing quantum technologies, it is possible to network quantum systems at a much larger scale in the near future. To deal with non-Markovian dynamics that is prevalent in solid-state devices, we propose a general transfer function based framework for modeling linear quantum networks, in which signal flow graphs are applied to characterize the network topology by flow of quantum signals. We define a noncommutative ring $\mathbb{D}$ and use its elements to construct Hamiltonians, transformations and transfer functions for both active and passive systems. The signal flow graph obtained for direct and indirect coherent quantum feedback systems clearly show the feedback loop via bidirectional signal flows. Importantly, the transfer function from input to output field is derived for non-Markovian quantum systems with colored inputs, from which the Markovian input-output relation can be easily obtained as a limiting case. Moreover, the transfer function possesses a symmetry structure that is analogous to the well-know scattering transformation in \sd picture. Finally, we show that these transfer functions can be integrated to build complex feedback networks via interconnections, serial products and feedback, which may include either direct or indirect coherent feedback loops, and transfer functions between quantum signal nodes can be calculated by the Riegle's matrix gain rule. The theory paves the way for modeling, analyzing and synthesizing non-Markovian linear quantum feedback networks in the frequency-domain.
\end{abstract}



\section{Introduction}
Under proper physical conditions (e.g., low temperature, ultrafast timescale), the emergence of quantum coherence effects provides new ways for information processing and precision \cite{Nielsen2000} that can break classical limitations. In recent years, this has been demonstrated to be possible by a large number of experimental breakthroughs in photonic, molecular and solid-state quantum devices \cite{Brif2010,You2011}, heralding the coming industrial applications of quantum technologies in the near future.

With advanced measurement and communication techniques, networking quantum devices is inevitable for large-scale quantum computation, communication and other applications. In such quantum networks, closed feedback loops may appear, intentionally or unintentionally, which complexifies the network topology. On the other hand, they bring opportunities from a control theoretic point of view \cite{Dong2010,Daless2008,Wu2012,Altafini2012}, as careful design of feedback loops may improve the precision and robustness of quantum systems in optical \cite{Wiseman2010,Mabuchi2011,Serafini2012,Crisafulli2013,Sayrin2011,Vinante2013,Shankar2013}.

Quantum feedback can be categorized into measurement-based feedback \cite{Jacobs2007} and coherent feedback \cite{Lloyd2000} strategies, depending on whether the feedback signal is classical (obtained by measurements) or quantum. Coherent feedback is further classified as direct or indirect strategies, which is analogous to the flyball governor \cite{Golnaraghi2010} or fly-by-wire flight control \cite{Stengel1993} in classical control engineering. The theory of coherent feedback control has been well developed based on an $(S,L,H)$ model \cite{Gough2009} driven by quantum white noises \cite{Gardiner,Hudson1984,Gardiner1985,Gough2008,Gough2010,Zhang2012b}, which can be conveniently used to describe interconnections of Markovian quantum systems \cite{Gough2009}. In particular, when the quantum network consists of only linear components \cite{Yanagisawa2003a,Yanagisawa2003b,James2008,Gough2008}, many modern control design methods, such as $H^\infty$ control \cite{James2008}, LQG control \cite{Nurdin2009} and model reduction \cite{Nurdin2013}, can be applied with an essential distinction on the physical realizability and treatment of noncommutative quantum noises.

Beside Markovian quantum networks, quantum networks involving non-Markovian components are drawing increasing attention. This is not only because Markovian approximation fails in most solid-state systems \cite{Clerk2010}, but also because non-Markovian phenomena is of great interests in theoretical physics \cite{Chruf2014}. In addition, properly engineered non-Markovian dynamics can be used for quantum optimal control and decoherence suppression \cite{Xue2012,Zhang2013}. In the literature, the characterization of non-Markovian quantum dynamics is mostly based on master equations with time-dependent coefficients or integral terms \cite{Gardiner1987,Diosi2012,Zhang2013}. In this paper, we are going to use transfer functions for this purpose, because it reduces the complexity from differential equations to algebraic equations, and frequency-domain analysis can be carried out based on experimental data obtained from spectral analyzers. There have been several studies in the literature for Markovian system \cite{Yanagisawa2003a,Yanagisawa2003b,Shaiju2012}, however, not much attention has been paid to quantum networks containing non-Markovian components with colored quantum noise inputs \cite{Zhang2013a,Wu2012a,Wu2013}.

The transfer function model to be established in this paper can be applied to both direct and indirect coherent quantum feedback systems, based on which signal flow graph method \cite{Mason1960,Riegle1972} will be introduced for graphical description of network topology and for network analysis. Although the signal flow graph approach is equivalent to the more frequently used block diagram representation, its advantageous lie in that the transfer functions between arbitrary source and sink nodes can be expressed in terms of path gains and loop gains read from the graph. This is well-known as Mason's or Riegle's gain rule \cite{Mason1960,Riegle1972}, which is convenient for systematic calculation in networks with overlapping feedback loops. Signal flow graphs have been broadly applied to the analysis of complex electrical, mechanical or hydrodynamic networks, but to our knowledge, no applications have been found in quantum networks.

The structure of this paper is as follows. Section \ref{Sec:Quantum Signals} will summarize basic concepts of quantum signals and introduce a quaternion-like ring for matrix representation of linear quantum dynamics. Section \ref{Subsec:DirectFeedback} shows how a signal flow graph can be constructed for direct feedback between interconnected quantum systems. Section \ref{Sec:InFeedback} derives the input-output relation for field-mediated indirect feedback systems, from which the Markovian limit can be easily obtained. Section \ref{Sec:Network} introduces basic components and connections for building quantum networks, following which a simple example of non-Markovian feedback network is provided and analyzed for demonstration. Finally, conclusions are drawn in Section \ref{Sec.Conclusion}.

\section{Preliminaries on Quantum dynamics and signal flow graphs}\label{Sec:Quantum Signals}
In this section, we will review the description of quantum signals and dynamics in linear quantum systems, from which a noncommutative quaternion-like ring $\HH$ is introduced for description of double-up vectors and transformations on them. The signal flow graph will also be reviewed in comparison with block diagram representation of control systems.

\subsection{Quantum signals}
A linear quantum system can be either bosonic (e.g., photons or Cooper pairs in semiconductors) or fermionic (e.g., electrons) obeying certain quantum statistical properties. A bosonic mode can be occupied by an arbitrary number of identical particles (e.g., photons). Let $|n\rangle_k$ be the number state with $n$ particles in the $k$-th mode, then all such states form an infinite-dimensional Hilbert space
$$\mathcal{H}_{\rm boson}=\bigotimes_{k=1,2,\cdots}{\rm span}\{|0\rangle_k,|1\rangle_k,\cdots\}$$
on which the system's operators are defined. The annihilation and creation operators satisfy
\begin{equation}\label{Eq:bosonic mode}
  [ \hat a_k, \hat a_{k'}^\dag]=\delta_{kk'},\quad   [ \hat a_k, \hat a_{k'}]=  [ \hat a_k^\dag, \hat a_{k'}^\dag]=0, \quad \forall k,k',
\end{equation}
where $[\hat  x, \hat y]=\hat  x \hat y- \hat y \hat x$ denotes the commutator.

In contrast, a fermionic mode can be occupied by at most one particle \cite{Dirac1982} and is ruled by Pauli exclusion principle. The Hilbert space of a $d$-mode fermion system is isomorphic to $\mathbb{C}^{2^d}$
and the annihilation and creation operators satisfy:
\begin{equation}\label{Eq:fermionic mode}
  \{ \hat a_k, \hat a_{k'}^\dag\}=\delta_{kk'},\quad   \{ \hat a_k, \hat a_{k'}\}=  \{ \hat a_k^\dag, \hat a_{k'}^\dag\}=0, \quad \forall k,k',
\end{equation}
where $\{\hat  x, \hat y\}=\hat  x \hat y+ \hat y \hat x$ denotes the anti-commutator.

\begin{remark}
The complete description of a quantum system also requires its density function $\rho$ (defined as a nonnegative definite and unit-trace Hermitian operator on the Hilbert space). Statistical properties of an arbitrary system observable are obtained by averaging over the density function, i.e., the expectation values of any function $f(\xx)$ can be thus calculated as $\langle f(\xx)\rangle=\tr(\rho f(\xx))$. For example, the quantum vacuum noise is defined under the vacuum state. The results obtained in this paper can be used with any quantum state, but we will not specify it as our focus is on the signal transfer properties in the Heisenberg picture.
\end{remark}

A quantum signal is referred to as a quantum observable varying in time or space, which conveys quantum information about the underlying physical system. In particular, suppose a quantum signal $\hat s(t)$ is a linear function of the system's annihilation and creation operators, say $\hat a_1(t),\cdots,\hat a_n(t)$, in the Heisenberg picture. Then we can always write
\begin{eqnarray*}
  \hat {s}(t) &=& \sum_{j=1}^n \left[\alpha_j \hat a_j(t)+\beta_j \hat a^\dag_j(t)\right], \\
  \hat {s}^\dag(t) &=& \sum_{j=1}^n \left[\beta_j^* \hat a_j(t)+\alpha_j^* \hat a^\dag_j(t)\right],
\end{eqnarray*}
where $\alpha_j$ and $\beta_j$ are arbitrary complex numbers. The essential difference of operator-valued quantum signals with classical signals is that they are noncommutative at different spacetime points.

We introduce a more compact expression by the following double-up operators:
\begin{equation}\label{}
 \hat{\bf a}=\left(\begin{array}{c}
                              \hat a \\
                              \hat a^\dag \\
                            \end{array}
                          \right)
\end{equation}
for an arbitrary mode $\hat a$, with which we have
$$\hat{\bf s}(t)=\sum_j \pp_j \aaa_j(t),$$
where the $2\times 2$ coefficient matrices
$$\pp_j= \left(
           \begin{array}{cc}
             \alpha_j & \beta_j  \\
             \beta_j^* &  \alpha_j^* \\
           \end{array}
         \right)\triangleq \Delta(\alpha,\beta).$$

Later we will see that such $2\times 2$ matrices can be used as ``numbers" to construct matrix transformations and $s$-functions. The collection of all such matrices form a noncommutative ring $\HH$ under the standard matrix sum and product operations. At the first glance, $\HH$ is very similar to the quaternion field
$$\mathbb{H}=\left\{ \left(
           \begin{array}{cc}
             \alpha & \beta \\
             -\beta^* &  \alpha^*\\
           \end{array}
         \right),\quad \alpha,\beta\in\mathbb{C}\right\}.
$$
However, they are algebraically different because that $(\HH\setminus\{0\},\cdot)$ does not form a group. Moreover, the subset of elements in $\HH$ with unit determinant form a noncompact group ${\rm SU}(1,1)$, while those in $\mathbb{H}$ form a compact group ${\rm SU}(2)$. In quantum physics, quaternion numbers can be used to define new quantum probabilities in the so called quaternionic quantum mechanics that are not completely equivalent to standard quantum mechanics \cite{Adler1995,Zhang1997}. Note that the $\HH$ is still rooted in standard quantum mechanics, and we use it for convenience of modeling and analysis of linear quantum networks.

Borrowing the notations of quaternion field, any $\pp\in\HH$ can be expanded as $\pp=a\ee+b\ii+c\jj+d\kk$, where $a,b,c,d\in\mathbb{R}$, under the following basis:
\begin{eqnarray*}
 \ee=\Delta(1,0),   && \ii=\Delta(i,0), \\
\jj=\Delta(0,1),   &&  \kk=\Delta(0,i).
\end{eqnarray*}
It can be verified that
$$-\ii^2=\jj^2=\kk^2=\ee,\quad\ii\jj\kk=\ee,$$
and
$$\ii\jj=-\jj\ii,\quad\jj\kk=-\kk\jj,\quad\kk\ii=-\ii\kk.$$

Now we can construct $\HH$ matrices with entries being $\HH$ numbers. Denote by $\HH^{m\times n}$ the set of $m\times n$ $\HH$ matrices. Similar to complex matrices, one can also define matrix rank, eigenvalues (as $\HH$ numbers) and eigenvectors (as $\HH$ vectors) for $\HH$ matrices (see appendix for details).

Another important generalization is the conjugate operation. Parallel with the complex conjugation of $\mathbb{C}$-numbers, the $\flat$-conjugate of a $\HH$-number is defined as follows:
$$\pp^\flat =\ii^{-1} \pp^\dag \ii,$$
where $\dag$ denotes the standard Hermitian conjugation. The $\flat$ operation changes $\pp=a\ee+b\ii+c\jj+d\kk$ to $\pp^\flat=a\ee-b\ii-c\jj-d\kk$. $\pp$ is said to be purely imaginary if $a=0$.
Similarly, the $\flat$ operation on a $\HH$-matrix $\mathbf{A}\in{\HH^{m\times n}}$ is defined as
$$\mathbf{A}^\flat = \JJ_m^{-1}\mathbf{A}^\dag \JJ_n,$$
where $\JJ_k={\rm diag}(\ii,\cdots,\ii)\in \HH^{k\times k}$.

A square $\HH$-matrix $\PP\in\HH^{n\times n}$ is said to be $\flat$-Hermitian (or skew $\flat$-Hermitian) if $\PP^\flat=\PP$ (or $\PP^\flat=-\PP$). An $\HH$-matrix $\mathbf{S}\in\HH^{n\times n}$ is called $\flat$-unitary (also known as an Bogoliubov transformation) if $\mathbf{S}^\flat \mathbf{S}=\mathbf{I}_n$, where $\mathbf{I}_n={\rm diag}\{\ee,\cdots,\ee\}$ is the identity matrix in $\HH^{n\times n}$.

The collection of $n\times n$ $\flat$-unitary matrices form a complex symplectic group ${\rm Sp}(2n,\mathbb{C})$, whose Lie algebra is the collection of skew $\flat$-Hermitian matrices. Since ${\rm Sp}(2n,\mathbb{C})$ has many similar properties with unitary group ${\bf U}(n,\mathbb{C})$, we also denote it by ${\bf U}(n,\HH)$. For example, the right eigenvalues of a skew $\flat$-Hermitian matrix must be purely imaginary, as well as skew Hermitian complex matrices. Such resemblance will greatly simplify the notations and analyses, which facilitates the understanding of linear quantum system structures.

\subsection{Linear quantum system dynamics}
The dynamics of a closed linear quantum system is governed by a quadratic Hamiltonian. For convenience, we always assume that the linear quantum systems (either bosonic or fermionic) studied in this paper contain a finite number of modes, but the results obtained can be extended to cases with enumerable or denumerable number of modes upon proper assumptions on convergence.
Suppose that the system contains $n$ modes with annihilation operators $\hat a_1,\cdots,\hat a_n$ that obey the above bosonic or fermion commutation relation, then the Hamiltonian must be in the following form
\begin{equation}\label{Eq:Hamiltonian}
 H={i}\sum_{1\leq k\leq j\leq n} \left(\alpha_{kj}\hat{a}_k^\dag \hat{a}_j-\alpha_{kj}^*\hat{a}_j^\dag \hat{a}_k+\beta_{kj}\hat{a}_k^\dag \hat{a}_j^\dag-\beta_{kj}^*\hat{a}_k \hat{a}_j \right)
\end{equation}
to guarantee its hermitian property, where $\alpha_{ij}$ and $\beta_{ij}$ are arbitrary complex numbers. The Hamiltonian generates a unitary transformation $U(t)=e^{-iHt}$ on any observable of the system, e.g., $\hat{a}_k(t)=U^\dag(t)\hat{a}_kU(t)$.

Using the following commutation relations
\begin{equation}\label{Eq:fermionic mode}
  [ \hat a_k^\dag \hat a_k, \hat a_{k'}]= -\delta_{kk'}\hat a_k,\quad  [\hat  a_k^\dag \hat a_k, \hat a_{k'}^\dag]= \delta_{kk'}\hat a_k^{\dag},
\end{equation}
which apply to both bosonic and fermionic systems, we can obtain the following Heisenberg equation of motion for $\aaa_j$:
\begin{equation}\label{Eq:Heisenberg Eq}
  \dot \aaa_j(t)= -i\left[ \aaa_j(t),H(t)\right]=\sum_{k=1}^n \pp_{kj}\aaa_j(t),
\end{equation}
for $j=1,\cdots,n$, where
$$\pp_{kj}=\left\{\begin{array}{ll}
                           \Delta(\alpha_{kj},\beta_{kj}), & k\leq j \\
                           \Delta(\alpha_{kj}^*,-\beta_{kj}), & k>j
                         \end{array}
\right.$$
are elements in $\HH$.

Denote the (first-order) state vector  as
$$\xx=\left(
        \begin{array}{c}
          \aaa_1 \\
          \vdots \\
          \aaa_n \\
        \end{array}
      \right),
$$
the overall evolution of the quantum system can then be written in a vector form
\begin{equation}\label{Eq:xdot=Px}
  \dot{\xx}(t) = \mathbf{P} \xx(t),
\end{equation}
where $\mathbf{P}= \{\pp_{jk}\}_{1\leq j,k\leq n}\in\HH^{n\times n}$ is skew $\flat$-Hermitian. Equation~(\ref{Eq:xdot=Px}) is formally similar to a \sd equation
$$\dot{\psi}(t)=-iH\psi(t)$$
in that the Hamiltonians $\PP$ and $-iH$ are both skew-hermitian under respective conjugate operations. Correspondingly, $\xx(t)$ follows $\flat$-unitary evolution $\xx(t)=e^{\PP t}\xx(0)$, as well as the unitary transformation over $\psi(t)$ under standard Hermitian conjugation.


\begin{remark}
In many existing studies, the state vector $\xx$ has entries arranged as follows \cite{Gough2010}
$$\xx=\left(\hat a_1,\cdots,\hat a_n; \hat a_1^\dag,\cdots,\hat a_n^\dag\right)^\top,$$
under which the coefficient matrix $\PP$ is in the following block form
\begin{equation}\label{Eq:A+-}
  \PP=\Delta(A_-,A_+)=\left(
        \begin{array}{cc}
          A_- & A_+ \\
          A_+^* & A_-^* \\
        \end{array}
      \right),
\end{equation}
where $A_\pm\in\mathbb{C}^{n\times n}$ and $A_\pm^*$ are the complex conjugate of $A_\pm$. This is equivalent to the $\HH$-matrix representation up to a permutation transformation. We choose to use the $\HH$-number based formalism because it is easier to understand the the dynamics by associating $n\times n$ (instead of $2n\times 2n$ matrices with an $n$-mode linear quantum system, and it is physically clear to encode the elementary properties (e.g., loss, gain and squeezing as summarized in Appendix \ref{Appendix:H algebra}) of each mode into a single $\HH$ eigenvalue. Moreover, as will be seen later, the similarities of $\flat$-unitary matrices with complex unitary matrices will facilitate the signal flow analyses in term of transfer functions.
\end{remark}

Finally, it should be noted that when the Hamiltonian (\ref{Eq:Hamiltonian}) includes only terms like $\hat a_i^\dag \hat a_j$ or $\hat a_i \hat a_j^\dag$ (see examples in Sections \ref{Subsec:DirectFeedback} and \ref{Sec:InFeedback}), each entry of $\PP$ must be in $2\times 2$ diagonal matrices, which form a commutative sub-ring $\HH_0=\{\Delta(\alpha,0),~\alpha\in\mathbb{C}\}$ of $\HH$. In such case, there is no squeezing effects, resulting in $A_+=0$ in (\ref{Eq:A+-}) and it is sufficient to consider only dynamics of $\hat a_j$'s governed by the $n\times n$ complex matrix $A_-$.

\subsection{Block diagrams and signal flow graphs}
\begin{figure}
\begin{center}
  \includegraphics[width=3.5in]{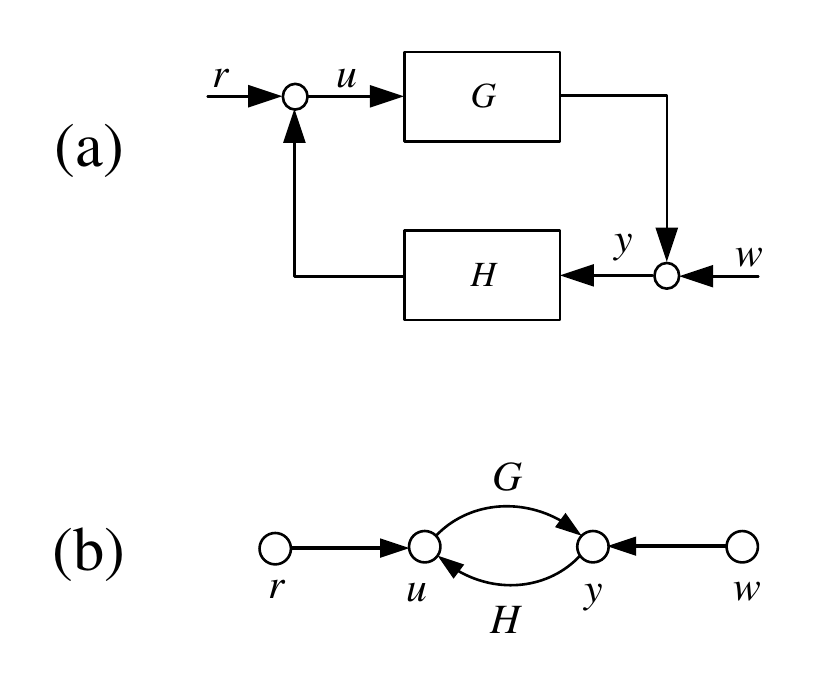}\\
    \caption{A feedback control system driven by a reference $r$ and a disturbance $w$, where $G$ and $H$ are the transfer functions of the plant and the controller: (a) the block diagram description; (b) the signal flow graph.}\label{Fig:SmallGain}
\end{center}
\end{figure}

In control theory, block diagram representation is often used to describe how the control acts on the system and how the components are connected. As shown in Fig.~\ref{Fig:SmallGain}(a), the signals are represented by directed arcs and the system is represented by a block associated with a transfer function. By contrast, a signal flow graph uses nodes to represent signals, while the system is represented by a directed arc. The directed arcs indicate the flow of information or energy between signal nodes, whose associated transfer functions describe how strong the originating nodes affect the terminating nodes. For simplicity, unit transfer gains will not be labeled in signal flow graphs. A node connected with only originating (terminating) arcs is called a source (sink) node. If a node is connected with multiple terminating arcs, then its value is equal to the sum of originating node values times the corresponding transfer gains.

Take the feedback control system with a disturbance as an example (see Fig.~\ref{Fig:SmallGain}(a)), which model is often used in robust control theory (e.g., small gain theorem). The control system is driven by a reference input $r$ against a disturbance $w$ that is added to the output signal $y$. The transfer functions of the plant and the controller are $G$ and $H$, respectively. The corresponding signal flow graph is shown in Fig.~\ref{Fig:SmallGain}(b), where the balance of signals read $u=r+Hy$ at node $u$ and $y=w+Gu$ at node $y$.

Block diagram representation is completely equivalent with signal flow graphs, but it is easier to understand for its resemblance with real systems. The advantage of signal flow graph is that the transfer gain between two arbitrary nodes can be systematically calculated according to the famous Mason's gain rule. In this paper, we take the advantages of signal flow graphs in conciseness and easiness of calculation from loop and path gains, and, to deal with matrix transfer function, we introduce Riegle's gain rule (see appendix for details) to calculate transfer functions between signal nodes in non-Markovian quantum networks.

\section{Signal flow in direction quantum feedback systems}\label{Subsec:DirectFeedback}
This section will derive the transfer function description of direct coherent feedback systems and its corresponding signal flow graphs. An example will be provided for demonstration.

\subsection{Direct quantum feedback systems}
Consider a quantum control plant described by Eq.~(\ref{Eq:xdot=Px}), and it is coupled to a direct coherent feedback controller implemented by an $m$-node linear quantum system with state vector:
\begin{equation}
  {\ww} = \left(
            \begin{array}{c}
              \bb_1 \\
              \vdots \\
              \bb_m \\
            \end{array}
          \right).
\end{equation}

The quantum system acts as a direct quantum feedback controller, with which the joint evolution must be in the following form:
\begin{equation}\label{Eq:Joint System}
     {\left(\!\!
                         \begin{array}{c}
                           \dot{\xx}(t) \\
                           \dot{\ww}(t)
                         \end{array}
                      \!\! \right)
     } =\left(\!\!
               \begin{array}{cc}
                 \mathbf{P} & \CC_{12} \\
                 \CC_{21} & \mathbf{Q}
               \end{array}
             \!\!\right)\left(\!\!
                         \begin{array}{c}
                         {\xx}(t) \\
                         {\ww}(t)
                         \end{array}\!\!
                       \right),
\end{equation}
where the coefficient matrices $\mathbf{P}\in\HH^{n\times n}$, $\mathbf{Q}\in\HH^{m\times m}$, $\CC_{12}\in\HH^{n\times m}$ and $\CC_{21}\in\HH^{m\times n}$.

Because the composite system is a closed linear quantum system, the overall coefficient matrix must be skew $\flat$-Hermitian, requiring that
$$\mathbf{P}^\flat=-\mathbf{P},\quad\mathbf{Q}^\flat=-\mathbf{Q},\quad\CC_{12}=-\CC_{21}^\flat.$$
Therefore, we can always write
\begin{equation}\label{Eq:NMLS1}
     {\left(
                         \begin{array}{c}
                           \dot{\xx}(t) \\
                           \dot{\ww}(t)
                         \end{array}
                       \right)
     } =\left(
               \begin{array}{cc}
                 \mathbf{P} & -\CC^\flat \\
                 \CC & \mathbf{Q}
               \end{array}
             \right)\left(
                         \begin{array}{c}
                         {\xx}(t) \\
                         {\ww}(t)
                         \end{array}
                       \right),
\end{equation}
where $\CC$ stands for the interaction between the two subsystems.

Performing Laplace transform on both sides of Eq.~(\ref{Eq:NMLS1}), we get
\begin{eqnarray}
\label{Eq:QCS-1}    {\xx}(s)&=&{\xx}_0(s)+\GG_x^w(s){\ww}(s),\\
\label{Eq:QCS-1'}   {\ww}(s)&=&{\ww}_0(s)+\GG_w^x(s){\xx}(s),
\end{eqnarray}
where quantum signals
$$\xx_0(s)=(s\mathbf{I}_n-\mathbf{P})^{-1}\xx(0),\quad {\ww}_0(s)=(s\mathbf{I}_m-\mathbf{Q})^{-1}{\ww}(0)$$
are system states in absence of interaction.
The transfer functions from $\ww(s)$ to $\xx(s)$ and from $\xx(s)$ to $\ww(s)$, respectively, are
$$\GG_x^{ w}(s)=(s\mathbf{I}_n-\mathbf{P})^{-1}\CC^\flat,~~\GG_{w}^x(s)=-(s\mathbf{I}_m-\mathbf{Q})^{-1}\CC.$$
In the following, we will drop the argument ``$s$" for $s$-functions in the Laplace domain unless it is necessary.

\begin{figure}
\begin{center}
  \includegraphics[width=3.5in]{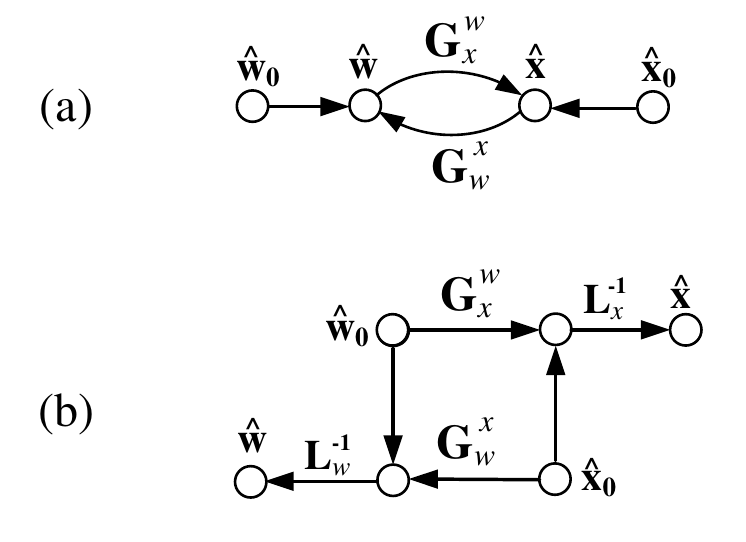}\\
    \caption{Signal flow graphs for a quantum system coupled to a direct coherent feedback controller: (a) the internal feedback loop induced by action and backaction; (2) the equivalent open-loop signal flow graph.}\label{Fig:Quantum Interaction}
\end{center}
\end{figure}

According to Eqs.~(\ref{Eq:QCS-1}) and (\ref{Eq:QCS-1'}), the signal flow graph between the two subsystems is depicted in Fig.~\ref{Fig:Quantum Interaction}(a). The bidirectional signal flows clearly show an internal feedback loop caused by ubiquitous action and backaction between the plant and the controller. By removing the feedback loop, as shown in Fig.~\ref{Fig:Quantum Interaction}(b), the net closed-loop transfer functions from $(\xx_0,\ww_0)$ to $(\xx,\ww)$ are
\begin{eqnarray}
\label{Eq:QCS-LT1}    \left(
                        \begin{array}{c}
                          \xx \\
                          \ww \\
                        \end{array}
                      \right)&=&     \left(
                        \begin{array}{cc}
                          \LL_x^{-1} & \LL_x^{-1}\GG_x^w \\
                          \LL_w^{-1}\GG_w^x & \LL_w^{-1} \\
                        \end{array}
                      \right)   \left(
                        \begin{array}{c}
                          \xx_0 \\
                          \ww_0 \\
                        \end{array}
                      \right),
\end{eqnarray}
where
\begin{eqnarray}
\LL_x=\mathbf{I}_n-\GG_x^{ w}\GG_{{ w}}^x,\quad \LL_{w}=\mathbf{I}_m-\GG_{ w}^x\GG_x^{ w}
\end{eqnarray}
are the loop differences (i.e., identity matrix minus the loop gain) of the feedback loop at $\xx$ and $\ww$, respectively. The feedback alters the transfer gain from $\xx_0$ to $\xx$ by the loop gain $\GG_x^\w\GG_w^x$, which can be used for preserving coherence dynamics, as will be shown in the following example.

\subsection{Example}
\begin{figure}
\begin{center}
  \includegraphics[width=3.5in]{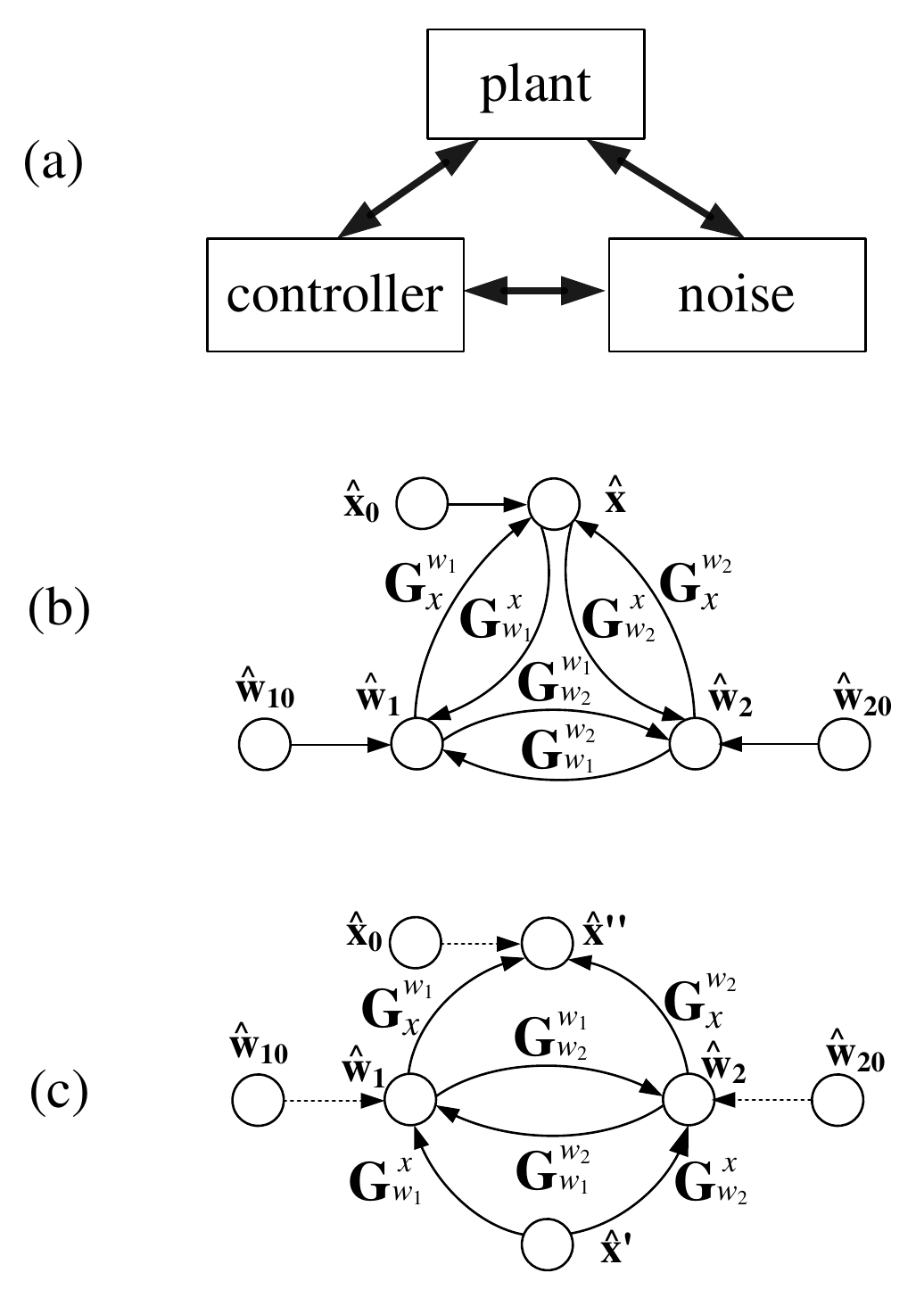}\\
    \caption{The signal flow graph for direct coherent feedback example: (a) the schematic setup for direct coherent feedback; (b) the signal flow graph under direct coherent feedback between controller and noise baths; (c) the split signal flow graph for calculating the FRL factor of $\xx$ on path $P_2$, where dotted arcs mean that the associated source nodes are not involved in the calculation.}\label{Fig:DirectFeedback}
\end{center}
\end{figure}

This example is taken from \cite{Xue2012}, but we will study it via the signal flow graph developed above. As shown in Fig.~\ref{Fig:DirectFeedback}(a), a single-mode quantum system to be controlled is bathed with two baths, which act as the noise and the controller, respectively. The joint Hamiltonian is given as follows:
\begin{eqnarray}
\nonumber  \hat H &=& \w_0\hat a^\dag \hat a +\sum_{k=1,2}\sum_{j=1}^m \w_{j}\hat b_{kj}^\dag \hat b_{kj}\\
&&+i\sum_{k=1,2}\sum_{j} \left(g_{kj}\hat a^\dag \hat b_{kj}-g_{kj}^*\hat a \hat b_{kj}^\dag\right),\label{Eq:H of Quantum Dot}
\end{eqnarray}
where $\hat a$, $\hat b_{1j}$'s and $\hat b_{2j'}$'s are, respectively, the annihilation operators of the plant, the controller bath and the noise bath. Correspondingly, the coefficient matrices read as
$$\PP=\w_0\ii, \quad \QQ_{1}=\QQ_2=\QQ={\rm diag}(\w_{1}\ii,\cdots,\w_{m}\ii),$$
and
$$\CC_k=\left(
          \begin{array}{c}
            \Delta(g_{k1},0) \\
            \vdots \\
            \Delta(g_{km},0)
          \end{array}
        \right), \quad k=1,2.
$$

The coherent feedback is introduced by directly coupling the controller bath with the noise bath, which modifies the Hamiltonian (\ref{Eq:H of Quantum Dot}) as
\begin{equation}
  H'=H+i\sum_k \left(f_k \hat{b}_{k1}\hat{b}^\dag_{k2}-f_k^* \hat{b}^\dag_{k1}\hat{b}_{k2}\right).
\end{equation}
Let $\mathbf{F}={\rm diag}\{\Delta(f_1,0),\cdots,\Delta(f_m,0)\}$, then
the resulting signal flow graph can be drawn as Fig.~\ref{Fig:DirectFeedback}(b), where
\begin{eqnarray*}
  \GG_x^{w_j}(s) &=& (s\ee-\PP)^{-1}\CC_j^\flat, \\
  \GG^x_{w_j}(s) &=& -(s\II_m-\QQ)^{-1}\CC_j, \\
  \GG_{w_1}^{w_2}(s) &=& (s\II_m-\QQ)^{-1}\mathbf{F}^\flat,\\
   \GG_{w_2}^{w_1}(s) &=& -(s\II_m-\QQ)^{-1}\mathbf{F}.
\end{eqnarray*}

The decoherence effect can be investigated by the transfer gain from $\xx(0)$ to $\xx$. When the noise and controller bath is not coupled, i.e.,  $\GG_{w_1}^{w_2}=\GG_{w_2}^{w_1}=0$, there are two feedback loops between the plant and the baths, and we derive that
\begin{eqnarray*}
 \xx  &=& \left(\ee-\GG_x^{w_2}\GG_{w_2}^x-\GG_x^{w_1}\GG_{w_1}^x\right)^{-1}\xx_{0},
\end{eqnarray*}
which, after substituting the parameters above, turns out to be
\begin{eqnarray*}
 \xx  &=& \left[s\ee-\ii \omega_0+\DD_{\rm free}(s)\right]\xx(0).
\end{eqnarray*}

Apparently, the system dynamics is differed from the original closed system dynamics by
$$\DD_{\rm free}(s)=\sum_k (|g_{k1}|^2+|g_{k2}|^2)(s\ee-\ii \omega_k)^{-1},$$
which can be used to quantify the decoherence effect.

When $\ww_1$ and $\ww_2$ are coupled together, a third feedback loop is introduced. According to Riegle's matrix gain rule, the total transfer gain from $\xx_0$ to $\xx$ is equal to the FRL factor of $\xx$. As shown in Fig.~\ref{Fig:DirectFeedback}(c), we split $\xx$ to calculate the FRL factor, in which there are two paths from $\xx'$ to $\xx''$ through $\ww_1$ and $\ww_2$, respectively, from which the loop difference is:
\begin{eqnarray*}
\LL_x&=&\ee-\GG_x^{w_1}(\II_{m}-\GG_{w_1}^{w_2}\GG_{w_2}^{w_1})^{-1}\GG_{w_1}^x\\
&&-\GG_x^{w_2}(\II_{m}-\GG_{w_2}^{w_1}\GG_{w_1}^{w_2})^{-1}\GG_{w_2}^x.
\end{eqnarray*}
Therefore, the closed-loop transfer function is
\begin{eqnarray*}
  \xx &=& \left[\ee-\GG_x^{w_1}(\II_{m_1}-\GG_{w_1}^{w_2}\GG_{w_2}^{w_1})^{-1}\GG_{w_1}^x\right.\\
  &&\left.-\GG_x^{w_2}(\II_{m_2}-\GG_{w_2}^{w_1}\GG_{w_1}^{w_2})^{-1}\GG_{w_2}^x\right]^{-1}\xx_{0}.
\end{eqnarray*}
Using the parameters above, we find
\begin{eqnarray*}
 \xx  &=& \left[s\ee-\ii \omega_0+\DD_{\rm fb}(s)\right]\xx(0),
\end{eqnarray*}
in which
$$\DD_{\rm fb}(s)=\sum_k (|g_{k1}|^2+|g_{k2}|^2)\left(s\ee-\ii \omega_k+\frac{|f_k|^2}{s\ee-\ii\omega_k}\right)^{-1}.$$
It can be seen that the decoherence part is altered by the direct feedback, by which one can properly choose $f_k$'s to modify the frequency response so as to suppress the decoherence effect near $\w_k\approx \w_0$. This provide a new angle to understand decoherence suppression strategy proposed in \cite{Xue2012}.

\section{Signal flows in field-mediated input-output quantum systems}\label{Sec:InFeedback}
In this section, we will define the inputs and outputs of non-Markovian quantum systems, based on which the transfer function from the input to the output is derived. The Markovian limit is then provided with connections to the $(S,L,H)$ model, and two demonstrative examples are given.

\subsection{Inputs and outputs of a quantum system coupled to a bath}
Equations~(\ref{Eq:QCS-1}) and (\ref{Eq:QCS-1'}) provides a general description for two directly interconnected quantum systems.
In practice, the subsystem $\ww$ may also act as an intermediate field that conveys information from one quantum system to another one. The intermediate field $\ww$ thus provides inputs and outputs, and by analyzing the relation between them one can extract information about the system.

Consider the model presented in Section \ref{Subsec:DirectFeedback}, where $\ww$ now represents the intermediate field. It is obvious that ${\ww}_0(t)$ should be the input field for it is only determined by the initial state of $\ww$ before interacting with $\xx$.

As for the output field, one may associate it $\ww(t)$. However, this is false because ``output" implies that the field should be free of interaction with $\xx$ after going through it, which is certainly not $\ww(t)$. The output field should be dynamically governed by only $\QQ$, and approaches to $\ww(t)$ after a sufficiently long time. In this regard, the output field should be defined as
$$\ww_{\infty}=\lim_{t_f\rightarrow \infty} e^{Q(t-t_f)}{\ww}(t_f),$$
where $\ww(t_f)$ is the final state at $t_f$.

Since each mode of $\xx$ is coupled to an observable of the intermediate field $\ww$ that is a linear combination of its modes (the coefficients correspond to a column of $\CC$), it is sufficient to study their input-output relations. In this way, we can reduce the dimension of the transfer function matrix, as $\ww$ usually contains a large (or even infinite) number of modes. Thus, corresponding to the $n$ columns of the matrix $\CC$, the system is coupled to at most $n$ effective fields or even less if these columns are linearly dependent (by Definition \ref{Def:Rank}). Let $k$ be the column rank of $\CC$ , then $\CC$ can be decomposed as
\begin{equation}\label{Eq:C=E*D'}
  \CC=\EE\cdot \DD^\flat,
\end{equation}
where $\EE\in \HH^{m\times k}$ represents the $k$ independent effective interaction channels and $\DD\in \HH^{n\times k}$ is their coupling matrix to the system.

\begin{figure}
\begin{center}
  \includegraphics[width=3.5in]{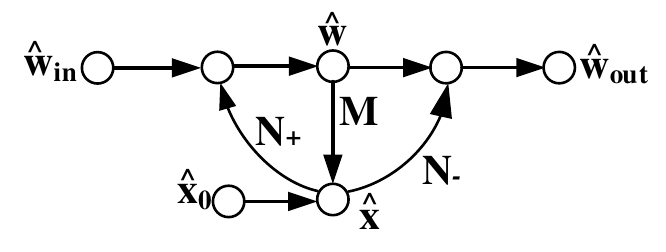}\\
    \caption{Signal flows from the input field $\ww_{\rm in}$ to the output field $\ww_{\rm out}$ through the quantum system $\xx$, where a feedback loop can be seen between nodes $\xx$ and $\ww$.}\label{Fig:Quantum IO2}
\end{center}
\end{figure}

Correspondingly, the effective input and output are defined as
$$\ww_{\rm in}(t)=\EE^\flat \ww_0(t),\quad \ww_{\rm out}(t)=\EE^\flat \ww_\infty(t),$$
which are both $k$-dimensional. In the following, we will derive the transfer function from $\ww_{\rm in}$ to $\ww_{\rm out}$.

\subsection{Transfer function description of the input-output relation}
Let $\NN(t)=\EE^\flat e^{\QQ t}\EE$ and $1(t)$ be the Heaviside step function. The following $s$-functions are defined as its Laplace transform integrated on negative and positive halves of the real axis:
\begin{eqnarray}
  \NN_\pm(s) &=& \int_{-\infty }^\infty \left[\NN(t)1(\pm t)\right]e^{-st}\dd t.
\end{eqnarray}
Similarly, we define $\MM(s)=\DD^\flat (s\II_n-\PP)^{-1}\DD$, which is associated with the system.

Our main conclusion is as follows:
\begin{theorem}\label{Th:1}
The transfer function from the input $\ww_{\rm in}(t)$ to the output $\ww_{\rm out}(t)$ is
$\GG(s)= \GG_-^{-1}(s)\GG_+(s)$, where $\GG_\pm(s)=\left[\II_k\pm\NN_\pm(s)\MM(s)\right]^{-1}$.
\end{theorem}

{\bf Proof:} According to (\ref{Eq:NMLS1}), we integrate the differential equation of $\ww$ from $t_0=0$ and $t_f=\infty$ to $t$, respectively,
\begin{eqnarray}
\label{Eq:QCS-2}   {\ww}_{\rm in}(t) & = & \EE^\flat{\ww}(t)+\int_{0}^t  \NN(t-\tau)\DD^\flat{\xx}(\tau)\dd \tau,\\
\label{Eq:QCS-2'}   {\ww}_{\rm out}(t) & = & \EE^\flat{\ww}(t)+\int_{\infty}^t  \NN(t-\tau)\DD^\flat{\xx}(\tau)\dd \tau,
\end{eqnarray}
from which $\ww(t)$ can be eliminated:
\begin{eqnarray}\label{Eq:QCS-3'}
{\ww}_{\rm out}(t) & =& {\ww}_{\rm in}(t)-\left(\int_{0}^t+\int_t^\infty\right) \NN(t-\tau)\DD^\flat[{\xx}(\tau)1(\tau)]\dd \tau.
\end{eqnarray}
Then, we perform Laplace transform on both sides of (\ref{Eq:QCS-3'} and obtain
\begin{equation}\label{Eq:IO with N+-}
\ww_{\rm out}(s)=\ww_{\rm in}(s)-
\left[\NN_+(s)+\NN_-(s)\right]\DD^\flat\xx(s).
\end{equation}

Then, we get the following differential-integral equation
\begin{eqnarray}
\label{Eq:QCS-3}   \dot{\xx}(t) & = & \PP{\xx}(t)-\int_{0}^t \DD \NN (t-\tau)\DD^\flat{\xx}(\tau)\dd \tau -\DD \ww_{\rm in}(t) \end{eqnarray}
by substituting (\ref{Eq:QCS-2}) into (\ref{Eq:NMLS1}), whose Laplace transform gives
\begin{equation}\label{Eq:IO with N+-}
  \DD^\flat \xx(s) =\left[\II_k+\MM(s)\NN_+(s)\right]^{-1}\left[\DD^\flat \xx_0(s)-\ww_{\rm in}(s)\right].
\end{equation}
Use this equation to replaced $\DD^\flat \xx$ in (\ref{Eq:IO with N+-}), and we have
\begin{eqnarray*}
  \GG(s) &=& \II_k- \left[\NN_+(s)+\NN_-(s)\right]\left[\II_k+\MM(s)\NN_+(s)\right]^{-1}\MM(s)\\
&=&\II_k- \left[\NN_+(s)+\NN_-(s)\right]\sum_{j=0}^\infty\left[-\MM(s)\NN_+(s)\right]^{j}\MM(s)\\
   &=& \II_k- \left[\NN_+(s)+\NN_-(s)\right]\MM(s)\left[\II_k+\NN_+(s)\MM(s)\right]^{-1}\\
   &=& \left[\II_k-\NN_-(s)\MM(s)\right]\left[\II_k+\NN_+(s)\MM(s)\right]^{-1},
\end{eqnarray*}
which ends of the proof.

Recall that in quantum scattering theory, the scattering transformation from the input to the output states (defined as ingoing and outgoing wavefunctions that are free of interactions with the scattering potential) has a similar form \cite{Dalton1999}:
\begin{equation}\label{Eq:Scattering}
  |\psi_{\rm out}\rangle=S|\psi_{\rm in}\rangle=\Omega_-^\dag \Omega_+|\psi_{\rm in}\rangle,
\end{equation}
where $\Omega_+$ and $\Omega_-$ are (unitary) M{\o}ller operators that connect the input and output states to the current state $|\psi\rangle$ of the system. Here, the transfer functions $\GG_\pm(s)$ play the same role as they represent the connections from $\ww_{\rm in}(t)$ and $\ww_{\rm out}(t)$ to the field $\ww(t)$ that is in interaction with the system.

From such an elegant analogy, it is natural to ask whether $\GG(s)=\GG_-^{-1}(s)\GG_+(s)$ has any unitary properties, because the scattering operator $S=\Omega_-^\dag\Omega_+$ is always unitary. We have the following conclusion:
\begin{theorem}\label{Th:2}
Denote $\GG^\sim(s)=\GG^\flat(-s^*)$. The transfer function satisfies $\GG^\sim(s)\GG(s)=\II_k$ as long as
$$[\NN_\pm(s),\MM(s)]=[\NN_+(s),\NN_-(s)]=0.$$
In particular, under this condition, $\GG(i\omega)$ is $\flat$-unitary, i.e., $\GG^\flat(i\omega)\GG(i\omega)=\II_k$.
\end{theorem}

{\bf Proof:} Using the following symmetries (see proof in Appendix \ref{Appendix:Symmetry}),
$$\MM^\sim(s)=-\MM(s),\quad \NN_\pm^\sim(s)=\NN_\mp(s),$$
we have
\begin{eqnarray*}
  \GG^\sim(s) &=& \left[\II_k-\MM(s)\NN_-(s)\right]^{-1}\left[\II_k+\MM(s)\NN_+(s)\right] \\
   &=& \left[\II_k-\NN_-(s)\MM(s)\right]^{-1}\left[\II_k+\NN_+(s)\MM(s)\right] \\
   &=&\left[\II_k+\NN_+(s)\MM(s)\right] \left[\II_k-\NN_-(s)\MM(s)\right]^{-1},
\end{eqnarray*}
which is equal to the inverse of $\GG(s)$. For the case of $s=i\omega$, $\GG^\sim(i\w)=\GG^\flat(i\w)$ and thus $\GG^\flat(i\w)\GG(i\w)=\II_k$. End of proof.

\begin{remark}
It is not known whether the condition in Theorem \ref{Th:2} is also necessary. On the other hand, since the decomposition $\CC=\EE\cdot\DD^\flat$ is nonunique, because $\EE'=\EE\mathbf{V}$ and $\DD'=\DD\mathbf{V}^{-\flat}$ for any $\mathbf{V}$ are valid for the decomposition. Therefore, the unitarity of the transfer function relies on the choice of effective inputs. It might be possible to find such a $\mathbf{V}$ matrix to turn a non-$\flat$-unitary transfer function to a $\flat$-unitary one, or in the opposite way.
\end{remark}

The property $\GG^\flat(i\omega)\GG(i\omega)=\II_k$ is known as all-pass property in passive quantum systems, which means $\GG(i\omega)$ changes only the phase but not the amplitude of the input field at all frequencies. It has been proven for Markovian systems in \cite{Shaiju2012} as is called the $(J,J)$ unitary property, which can be taken as a special case of Theorem \ref{Th:2}.  However, this unitary property may not hold for general non-Markovian systems (see examples below) when the assumptions are violated. This is an important distinction between Markovian and non-Markovian systems.

The input-output formalism can be generalized to systems coupled to multiple noninteracting fields, where the system $\xx$ can be taken as a coupler (or switch, router) that modulates the input-output relations between $\ww_1,\cdots,\ww_q$. Let $m_j$ be the number of modes contained in $\ww_j$, $j=1,\cdots,q$, then we can take them as a whole larger field corresponding to:
\begin{equation}\label{Eq:NMLSk}
     \QQ =\left(
               \begin{array}{cccc}
                 \QQ_{1} &  &   \\
                  &\ddots &  \\
                  & & \QQ_q
               \end{array}
             \right),
\end{equation}
where $\QQ_j\in\HH^{m_j\times m_j}$ and
$$\CC=\DD\cdot\EE^\flat = \left[\DD_1,\cdots,\DD_q\right]\cdot{\rm diag}(\EE_1\,\cdots,\EE_q).$$
Therefore, Theorem \ref{Th:1} can be applied with
\begin{eqnarray}
\label{Eq:MIMO-M} \MM(t)  &=& \left(
           \begin{array}{ccc}
             \DD_1^\flat e^{\PP t}\DD_1 & \cdots & \DD_1^\flat e^{\PP t}\DD_q \\
            \vdots  &\ddots  & \vdots \\
             \DD_q^\flat e^{\PP t}\DD_1 & \cdots & \DD_q^\flat e^{\PP t}\DD_q  \\
           \end{array}
         \right), \\
\label{Eq:MIMO-N}\NN(t)&=&\left(
           \begin{array}{ccc}
            \EE_1^\flat e^{\QQ t}\EE_1 &  &  \\
              &\ddots  &  \\
              &  & \EE_q^\flat e^{\QQ t}\EE_q  \\
           \end{array}
         \right).
\end{eqnarray}

\subsection{Markovianess and its Markovian limit}
First, we introduce the following adjoint $\flat$ operation
\begin{equation}\label{}
 \ww^\flat=(\bb_1^\flat,\cdots,\bb_m^\flat)
\end{equation}
where $\bb_k^\flat=(i\hat b_k^\dag , -i\hat b_k)$ for $k=1,\cdots,m$, for compatibility with $\HH$-matrix $\flat$ operations such that
$$\left(\mathbf{A}\mathbf{B}\right)^\flat=\mathbf{B}^\flat\mathbf{A}^\flat,\quad \left(\mathbf{A}\xx\right)^\flat=\xx^\flat\mathbf{A}^\flat$$
are valid for any proper operator vector $\ww$ and $\HH$ matrices $\mathbf{A}$ and $\mathbf{B}$.

Using this notation, it is easy to prove the fundamental commutation relation $[\ww,\ww^\flat]=i\II_m$, which is preserved during the evolution, i.e., $[\ww(t),\ww^\flat(t)]=[\ww(0),\ww^\flat(0)]=i\II_n$. Accordingly, the effective inputs and outputs must satisfy
\begin{equation}\label{Eq:ColorNoise}
\left[\ww_{\rm in}(t),\ww_{\rm in}^\flat(t')\right]=\left[\ww_{\rm out}(t),\ww_{\rm out}^\flat(t')\right]=\NN(t-t').
\end{equation}
because, for example,
\begin{eqnarray*}
\left[\ww_{\rm in}(t),\ww_{\rm in}^\flat(t')\right]
&=&\left[\EE^\flat e^{\QQ t}\ww(t_0),(\ww(t_0))^\flat e^{\QQ^\flat t'}\EE\right]
\\
&=&\EE^\flat e^{\QQ t}\left[\ww(t_0),(\ww(t_0))^\flat\right]e^{-\QQ t'}
\EE\\
&=&i\,\EE^\flat e^{\QQ (t-t')}
\EE.
\end{eqnarray*}
This implies that, when taking $\ww_{\rm in}(t)$ as an input noise, it is usually colored because the correlation time is finite. Its spectral properties is dependent on the Fourier transform of $\NN(t)$. On the other hand, owing to the nonsingular integral term in Eq.~(\ref{Eq:QCS-3}), such colored noise inputs lead to non-Markovian dynamics of the system $\xx$, and it is also the function $\NN(t)$ that determines non-Markovianity of the dynamics.

Therefore, we can can obtain the Markovian limit from the general non-Markovian system by pushing the correlation time to zero, i.e., $\NN(t)\rightarrow\NN_{0}\delta(t)$. In such case, $\NN_+(s)=\NN_-(s)=\frac{1}{2}\NN_0$ are constant $\flat$-Hermitian matrices. The resulting Markovian different equation of $\xx$ becomes
\begin{eqnarray}
\label{Eq:MarkovianLimit1}  \dot\xx(t) &=& \left(\PP-\frac{1}{2}\DD\NN_{0}\DD^\flat\right)\xx(t)+\DD{\ww}_{\rm in}(t),\\
\label{Eq:MarkovianLimit2}  \ww_{\rm out}(t) &=& \ww_{\rm in}(t)+\NN_{0}\DD^\flat\xx(t),
\end{eqnarray}
from which the transfer function from the effective input to the effective output is
$$\GG(s)=\left[\II_k-\frac{1}{2}\DD^\flat(s\II_n-\PP)^{-1}\DD\NN_0\right]\left[\II_k+\frac{1}{2}\DD^\flat(s\II_n-\PP)^{-1}\DD\NN_0\right]^{-1}.$$

Note that the perturbation $\frac{1}{2}\DD\NN_0\DD^\flat$ to $\PP$ may lead to either loss or gain. For example, when $\NN_0=\ee$, then loss is present when $\DD=\ee$. If $\DD=\kk$, then the intermediate field produces gain (e.g., through two-photon processes in quantum optics) in the system and causes instability.

The canonical quantum white noises (under proper states, e.g., vacuum state) that is broadly used in the literature \cite{Hudson1984} correspond to $\NN_{0}=\II_k$\cite{Zhang2012,Hudson1984}. In such case, it is easy to find the connection to the $(S,L,H)$ model via $\HH$ matrices by setting, respectively,
\begin{equation}\label{eq:SLH}
  S=\II_k,\quad L=\DD^\flat \xx(t),\quad H=\frac{1}{2}\xx^\flat(t)\PP\xx(t)
\end{equation}
as the scattering matrix, the system's coupling operator and the system's internal Hamiltonian. Moreover, verifying that the conditions of Theorem \ref{Th:2} are satisfied because $\NN_\pm(s)=\frac{1}{2}\II_k$, we immediately prove that the corresponding transfer function is $\flat$-unitary for $s=i\w$, as is proven in \cite{Shaiju2012}. Note that this may not be true when $\NN_0\neq \II_k$.

\subsection{Examples}
Consider a single-mode passive optical cavity bathed with a collection of bosonic modes. Under the rotating-wave approximation, the total Hamiltonian can be written as:
$$ H =\w_0a^\dag a+ \sum_k \omega_k \hat b^\dag_k  \hat b_k+i\sum_{k} \left(g_k\hat a \hat b^\dag_k- g_k^*\hat a^\dag \hat b_k\right),$$
corresponding to
$\mathbf{P}=-\ii\w_0$ and
$$\QQ=-\left(
        \begin{array}{ccc}
\w_1\ii &  &  \\
           & \ddots &  \\
           &  &\w_m\ii \\
        \end{array}
      \right),\quad\CC=\left(
               \begin{array}{c}
                 \Delta(g_1,0) \\
                 \vdots \\
                 \Delta(g_m,0) \\
               \end{array}
             \right),
$$
and the signal flow graph can be represented by Fig.~\ref{Fig:Quantum Interaction}. There is only one effective input and therefore we decompose $\CC=\EE\cdot \DD^\flat$, where $\EE=\CC$ and $\DD=\ee$. This corresponds to the input field
$$\ww_{\rm in}(t) =\sum_k \Delta(g_k,0)e^{-\ii \w_kt}\bb_k(0).$$

Therefore, we have
$$\NN(t)=\sum_k \Delta^\flat(g_k,0) e^{-\ii\w_k t}\Delta(g_k,0)=\sum_k |g_k|^2 e^{-\ii\w_k t}\rightarrow \int_{-\infty}^\infty |g(\w)|^2 e^{-\ii\w t}{\rm d}t.$$

For example, if we take the Lorentzian spectral shape $g(\w)=\sqrt{\frac{\kappa\gamma^2}{\w^2+\gamma^2}}$, then $\NN(t)=\frac{\kappa\gamma}{2}e^{-\gamma|t|}\ee$ and correspondingly,
$$\NN_+(s)=\frac{\kappa\gamma}{2}(s+\gamma)^{-1}\ee,\quad \NN_-(s)=\frac{\kappa\gamma}{2}(-s+\gamma)^{-1}\ee,$$
and $\MM(s)=(s\ee-\ii\w)^{-1}$.
The transfer function from the effective input to the effective output fields is
\begin{equation}\label{Eq:TF for example 2}
 \mathbb{\GG}(s) = \left[\ee +\frac{\kappa\gamma}{2}(s\ee-\w\ii)^{-1}(s-\gamma)^{-1}\right]\left[\ee+\frac{\kappa\gamma}{2}(s\ee-\w\ii)^{-1}(s+\gamma)^{-1}\right]^{-1}=\left(
                                                                                                      \begin{array}{cc}
                                                                                                        G_+(s) &  \\
                                                                                                         & G_-(s)\\
                                                                                                      \end{array}
                                                                                                    \right)
 ,
\end{equation}
where $$G_\pm(s)=\frac{s\mp i\w+\frac{\kappa\gamma}{2}(s-\gamma)^{-1}}{s\mp i\w+\frac{\kappa\gamma}{2}(s+\gamma)^{-1}}.$$

Now let us take the Markovian limit by pushing $\gamma\rightarrow \infty$, which leads to $\lim_{\gamma\rightarrow \infty} \NN_\pm(s)=\pm \frac{\kappa}{2}$, then the transfer function becomes
$$G_\pm(s)=\frac{s\mp i\w-\frac{\kappa}{2}}{s\mp i\w+\frac{\kappa}{2}},$$
which exactly recovers the results obtained in \cite{Yanagisawa2003a}.

Note that it is sufficient to characterize the transfer function merely by the scalar function $G_+(s)$. However, if the control plant is an active cavity (e.g., a degenerated parameter amplifier) with $\PP=\sigma\kk$, we have to use the full expression
\begin{equation}
 \mathbb{\GG}(s) = \left[s\ee-\sigma\kk+(s-\gamma)^{-1}\ee\right]\left[s\ee-\sigma\kk+(s+\gamma)^{-1}\ee\right]^{-1},
\end{equation}

Next, we show that $\GG(i\omega)$ is not always $\flat$-unitary if the condition in Theorem \ref{Th:2} is violated. Take the example in Section \ref{Subsec:DirectFeedback} for example. In absence of coupling between the two baths, we can set $\EE_k=\CC_k$ and $\DD_k=\ee$, $k=1,2$ and use (\ref{Eq:MIMO-M}) and (\ref{Eq:MIMO-N}) to get
$$\MM(s)=\left(
           \begin{array}{cc}
             (s\ee-\ii \w_0)^{-1} & (s\ee-\ii \w_0)^{-1} \\
             (s\ee-\ii \w_0)^{-1} & (s\ee-\ii \w_0)^{-1} \\
           \end{array}
         \right).
$$
and
$$\NN_\pm(s)=\frac{1}{2}\left(
           \begin{array}{cc}
             \kappa_1\gamma_1(\pm s+\gamma_1)^{-1}\ee &  \\
                 & \kappa_2\gamma_2(\pm s+\gamma_2)^{-1}\ee \\
           \end{array}
         \right),
$$
where Lorentzian spectrum is adopted with widths $\gamma_1$ and $\gamma_2$. Then we have
\begin{eqnarray*}
\GG&=&\left(\begin{array}{cc}
                      (s\ee-\ii \w_0)+(s-\gamma_1)^{-1}\ee & (s-\gamma_1)^{-1}\ee \\
                      (s-\gamma_2)^{-1}\ee & (s\ee-\ii \w_0)+(s-\gamma_2)^{-1}\ee \\
                    \end{array}
                  \right)\\
                  &&\cdot \left(\begin{array}{cc}
                      (s\ee-\ii \w_0)+(s+\gamma_1)^{-1}\ee & (s+\gamma_1)^{-1}\ee \\
                      (s+\gamma_2)^{-1}\ee & (s\ee-\ii \w_0)+(s+\gamma_2)^{-1}\ee \\
                    \end{array}
                  \right)^{-1}.
\end{eqnarray*}
According to Theorem \ref{Th:2}, the condition $[\NN_+(s),\NN_-(s)]=0$ is always satisfied, but the condition $[\NN_\pm(s),\MM(s)]=0$ holds only when $\gamma_1=\gamma_2$ and $\kappa_1=\kappa_2$. When $\gamma_1\neq \gamma_2$, one can verify that $\GG(i\omega)$ is not $\flat$-unitary.

\section{Non-Markovian Coherent feedback quantum Networks}\label{Sec:Network}
In this section, we will first introduce some basic components for networking quantum systems in terms of $\HH$ matrices and transfer functions, as well as the series product operation for cascading quantum systems. Then, via a simple example of coherent feedback system via field mediated interactions, we show how the signal flow graph is constructed and analyzed by Riegle's matrix gain rule.

\subsection{Beam splitter and time-delay in quantum networks}\label{Sec:Components}
\begin{figure}
\begin{center}
  \includegraphics[width=3.5in]{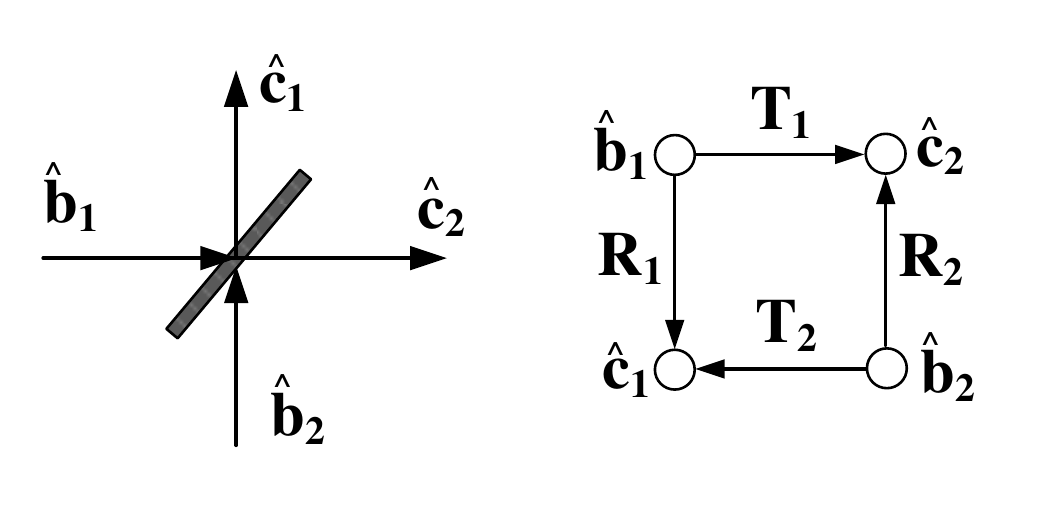}\\
    \caption{The signal flow graph for a beam splitter: (left) optical implementation by a partial transmission and partial reflection mirror; (right) the corresponding signal flow graph.}\label{Fig:BeamSplitter}
\end{center}
\end{figure}

Beam splitters \cite{Leonhardt2003} are used to mix separate quantum signals or split a quantum signal into different channels (see Fig.~\ref{Fig:BeamSplitter}). The two input and two output signals are related by a linear transformation
\begin{equation}\label{Eq:Beam Splitter}
  \left(
    \begin{array}{c}
       \cc_1 \\
       \cc_2 \\
    \end{array}
  \right)= \left(
             \begin{array}{cc}
               \mathbf{T}_1 & \mathbf{R}_1 \\
               \mathbf{R}_2 & \mathbf{T}_2\\
             \end{array}
           \right)
    \left(
    \begin{array}{c}
       \bb_1\\
       \bb_2\\
    \end{array}
  \right),
\end{equation}
where $\cc_1$ ($\bb_1$) and $\cc_2$ ($\bb_2$) contain $m_1$ and $m_2$ modes, respectively. The matrices $\mathbf{T}_1\in\HH^{m_1\times m_1}$ and $\mathbf{T}_2\in\HH^{m_2\times m_2}$ are the transmission matrix of all modes in the signal through the beam splitter; and $\mathbf{R}_1\in\HH^{m_1\times m_2}$ and $\mathbf{R}_2\in\HH^{m_2\times m_1}$ are the reflection matrix of all modes from the beam splitter. The entire $\HH$-matrix is a $\flat$-unitary matrix in ${\rm USp}(2n,\mathbb{C})$ so as to preserve the commutation relationship between the entries of the input and output fields. Moreover, because beam splitters are passive components, the entire $\HH$ matrix should also be unitary in the sense of complex-number conjugation.

Constant $\HH$ matrices can also be used to describe static components such as attenuators, amplifiers, spectral filters, multiplexers or demultiplexers in photonic systems \cite{Saleh2007}. Due to the limit of length, we will not discuss them here.

Another important element of quantum networks is the time-delay, which is inevitable in waveguides. In quantum optics, the time delay is induced by the propagation over a distance (e.g., long-distance optical fibers), and the amount of time delay is equal to the distance $L$ divided by the speed of propagation. For non-dispersive waveguides in which all modes have a uniform speed $c$, the time delay can be represented by
$$\mathbf{G}(s)=e^{-sL/c}\II_m,$$
where $m$ is the number of modes in the waveguide. However, in dispersive waveguides, the transfer function has to be written as the following diagonal matrix
$$\mathbf{G}(s)=\left(
                   \begin{array}{ccc}
                    e^{-sL/c_1}\ee  &  &  \\
                      & \ddots &  \\
                      &  & e^{-sL/c_m}\ee \\
                   \end{array}
                 \right),$$
where $c_k$, $k=1,\cdots,m$, is the speed of the $k$-th mode in the waveguide. Time delay is usually unwanted, but sometimes can be utilized for feedback-controlled lasers \cite{Ohtsubo2012}. In the following example, we will take it into account.

\subsection{Series product of linear quantum systems}\label{Subsec:Cascade}
When two quantum systems are not directly interacted, a travelling field can transfer interactions from one to the other. Such intermediate field cascades the two systems via series product operation. Each system is coupled to a field, and the output field of the first system is fed into the second system as an input field.

Note that because the signal flow from the first system to the second system is unidirectional, a nonreciprocal device (e.g., an isolator or a circulator) has to be applied like diodes in electrical circuits. Such devices cannot be modeled by a linear dynamical system as proposed in Section \ref{Subsec:DirectFeedback}, but it is easy to describe it by a signal flow graph, i.e., a unidirectional arrow from one node to another.

The total transfer function of cascaded systems via series product is easy to calculated. Suppose that the transfer functions of the two systems are $\GG_{\pm,j}=\left[\II_k\pm \NN_{\pm}(s)\MM_j(s)\right]^{-1}$, $j=1,2$, the total transfer function is simply their product:
\begin{eqnarray}
\GG(s) &=& \GG_{-,2}^{-1}(s)\GG_{+,2}(s)\GG_{-,1}^{-1}(s)\GG_{+,1}(s). \label{Eq:mode series product}
\end{eqnarray}

\subsection{Example: Indirect coherent feedback system with a single-input sytem}

\begin{figure}
\begin{center}
  \includegraphics[width=3.5in]{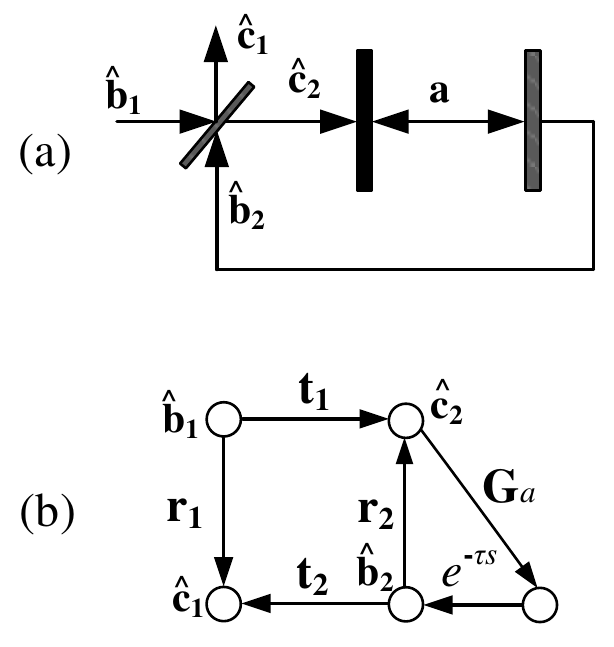}\\
    \caption{The signal flow graph for the example of indirect coherent feedback: (a) optical setup for the closed-loop control system; (b) the signal flow graph.}\label{Fig:Feedback1}
\end{center}
\end{figure}

This example is based on an example in \cite[Fig.~6]{Yanagisawa2003a}. As shown in Fig.~\ref{Fig:Feedback1}, the input field $\bb_1$ is fed through a beam splitter into a single-mode optical cavity, whose output field $\bb_2$ is directed back to the other input channel of the beam splitter, along which a time delay is present. The output field $\cc_1$ is at the other output port of the beam splitter. Next, we use the signal flow graph to calculate the closed-loop transfer function from $\bb_1$ to $\cc_2$.

The corresponding signal flow graph is shown in Fig.~\ref{Fig:Feedback1}(b). Now we apply the Riegle's gain rule (see Appendix for a summary) to the calculation of transfer function from $\bb_1$ to $\cc_2$. We find two paths as follows:
\begin{enumerate}
  \item $P_1=\bb_1\rightarrow \cc_1$,
  \item $P_2=\bb_1\rightarrow \cc_2\rightarrow\bb_2\rightarrow \cc_1$.
\end{enumerate}

The gain contributed by $P_1$ is $\GG_1=\mathbf{r}_1$ as there are no intermediate nodes. For $P_2$, the path gain is $\mathbf{t}_2 \GG_a\mathbf{t}_1$,
where $\GG_a(s)$ is derived in (\ref{Eq:TF for example 2}). Only the FRL factor $F_2(\bb_2)=\left(\ee-e^{-\tau s}\GG_a\mathbf{r}_{2}\right)^{-1}$ at $\bb_2$ affects the total transfer gain, and thereby we have
\begin{eqnarray*}
  \GG &=&\GG_1+\GG_2 \\
  &=& \mathbf{r}_{1}+\mathbf{t}_{2}\left(e^{\tau s}\GG^{-1}_a-\mathbf{r}_{2}\right)^{-1}\mathbf{t}_{1}\\
\end{eqnarray*}

As well as the control plant itself, the closed-loop transfer function is still all-pass because this property is not changed under a fractional transformation \cite{Shaiju2012,Zhang2012}. Thus, the feedback only affects the phase characteristics of the system. This feature may be used to identify non-Markovianity or noise spectrum from phases characteristics measured from an unknown system.

\section{Conclusion}\label{Sec.Conclusion}
To conclude, we presented a framework for modeling linear non-Markovian quantum networks by transfer functions and signal flow graphs, which is based on a noncommutative ring $\HH$ that has not been used before. The signal flow graph clearly shows closed loops appearing in direct and indirect coherent feedback systems. The resulting input-output transfer function for field-mediated systems nontrivially generalizes the Markovian case, and has an elegant analogy to the scattering transformation in \sd picture. Various examples were provided to show how the transfer functions and signal flow graphs are obtained in non-Markovian quantum networks, and how transfer gains are calculated via Riegle's gain formula.

This framework can also be applied to networks of classical oscillators, but in the classical domain there is no counterpart of networks of fermionic systems, as well as non-commutativity between in quantum observables. The transfer function based signal flow graphs provide a basis for frequency-domain modeling, analysis and synthesis over complex quantum networks, which embraces the non-Markovian nature of generic quantum systems.

In practice, the resulting transfer functions may become irrational for general field couplings, and thereby it could be very hard to extract the time-domain response from the frequency-domain expression. However, we indicate that under many circumstances the control performance can be directly evaluated by the frequency response (e.g., Nyquist and Bode plots) without having to know the time-domain solutions. Moreover, the framework can be extended to analyze the motion of higher-order moments and higher-order correlation properties that are more essential in quantum statistics. These topics will be studied in the future.

The framework presented here opens up many opportunities for studying control of non-Markovian quantum networks from a frequency-domain point of view.  We expect that it can be combined with QHDL (Quantum Hardware Description Language) in practical design and control of quantum networks \cite{Tezak2012}.

\appendix
\subsection{Algebraic properties of $\HH$}\label{Appendix:H algebra}
Similar to real or complex matrices, the concepts of rank, eigenvalues and eigenvectors can be introduced for analysis of $\HH$-matrix.
\begin{definition}\cite{Xu2003}\label{Def:Rank}
The vectors $\vec{\mathbf{v}}_1,\cdots,\vec{\mathbf{v}}_k\in \HH^{n\times1}$ are linearly independent if $\mathbf{c}_1\vec{\mathbf{v}}_1+\cdots+\mathbf{c}_k\vec{\mathbf{v}}_k=0$, where $\mathbf{c}_1,\cdots,\mathbf{c_k}\in\HH$, implies $\mathbf{c}_1=\cdots=\mathbf{c}_k={\bf 0}$. The row (or column) rank of a matrix $\mathbf{A}\in\HH^{n\times m}$ is referred to as the largest number of linearly independent rows (or columns) in $\mathbf{A}$.
\end{definition}

\begin{definition}\cite{Xu2003}
An $\HH$-number ${\bf h}$ is said to be a right eigenvalue of $\mathbf{A}\in\HH^{n\times n}$, if there exists a vector $\vec{\mathbf{v}}\in\HH^{n\times 1}$, such that ${\bf A} \vec{\mathbf{v}}=\vec{\mathbf{v}}\mathbf{h}$.
\end{definition}

Note that, different from a real or complex matrix,  an $\HH$-matrix has infinite right eigenvalues because if $\mathbf{h}$ is a right eigenvalue of $\mathbf{A}$, then $\mathbf{h}'=\mathbf{u}^\flat\mathbf{h}\mathbf{u}$ for any unimodular number $\mathbf{u}$ (i.e., $\mathbf{u}^\flat\mathbf{u}=\ee$) is also a right eigenvalue because
$$\mathbf{A} \vec{\mathbf{v}}\mathbf{u}=\vec{\mathbf{v}}\mathbf{u}\left(\mathbf{u}^\flat\mathbf{h}\mathbf{u}\right).$$

Suppose that ${\bf h} = a\ee+b\ii+c\jj+d\kk$ is a right eigenvalue of ${\bf P}$, then each class can be represented by $a$ and $C=-b^2+c^2+d^2$ that are invariant under unimodular transformations. They can be used to characterize optical properties of an electromagnetic mode in waveguides:
\begin{enumerate}
  \item $a>0$ ($a<0$) implies that the mode is in a gain (lossy) medium;
  \item $C>0$ implies that the mode is squeezed, otherwise it is not squeezed.
\end{enumerate}

\subsection{Riegle's matrix gain rule}\label{Appendix:GainRule}
Consider the transfer function from a source node $A$ to a sink node $B$ in a linear quantum network. The Riegle's gain rule is stated as follows \cite{Riegle1972}:
\begin{enumerate}
  \item Find out all forward paths that have no self-intersections from $A$ to $B$, say $P_1,~P_2, \cdots,~P_k$;
  \item The contribution $\GG_j$ of a path ${P_j}$ to the total transfer gain $\GG$ is equal to the path gain of ${P}_j$ interrupted by the forward return loop (FRL) factors (to be explained below). For example, suppose that the path$$P_j=A\rightarrow C_1\rightarrow C_{2}\rightarrow B$$ contains two intermediate nodes $C_1$ and $C_2$, and its path gain is $\GG_{B}^{C_2}\GG_{C_2}^{C_1}\GG_{C_1}^A$. Let $F_j(C_1)$ and $F_j(C_2)$ be the FRL factors of $C_1$ and $C_2$, respectively, then the contribution of $P_j$ is $$\GG_j=\GG_{B}^{C_2}\cdot\FF_j(C_2)\cdot\GG_{C_2}^{C_1}\cdot\FF_j(C_1)\cdot\GG_{C_1}^A.$$
 \item The total transfer gain from $A$ to $B$ is the sum of the contributions of each path as given in step 2, i.e., $$\GG=\GG_1+\cdots \GG_k.$$
\end{enumerate}

To calculate the FRL factor of a node $C$ on a path $P_j$ from $A$ (the source node) to $B$ (the sink node), we first separate it from all nodes on the path between $C$ and $B$. Then, we split $C$ into a source node $C'$ connected with all outgoing arcs and a sink node $C''$ connected with all ingoing arcs. The loop difference of $C$ is defined as the identity matrix minus the transfer gain from $C'$ to $C''$. The FRL factor $F_j(C)$ of $C$ on the path $P_j$ is the inverse of its loop difference of this node.

\subsection{Symmetries in $\NN_\pm(s)$ and $\MM(s)$}\label{Appendix:Symmetry}
\subsubsection{Proof of $\MM^\sim(s)=-\MM(s)$}
$$\MM^\sim(s)=\DD^\flat((-s^*)^*\II-\PP^\flat)^{-1}\DD=-\MM(s).$$

\subsubsection{Proof of $\NN_\pm^\sim(s)=\NN_\mp(s)$}
According to definition $\NN(t)=\EE^\flat e^{\QQ t}\EE$, we have $\NN^\flat(t)=\EE^\flat e^{\QQ^\flat t}\EE=\NN(-t)$, showing that it is an even function of time. In addition, denote $\mathbf{X}^\sim(s)=\mathbf{X}^\flat(-s^*)$, then
$$\NN^\sim_-(s)=\int_{-\infty}^0 \NN^\flat(t)e^{-(-s^*)^*t}\dd t=\int_{-\infty}^0 \NN(-t)e^{st}\dd t=\int_0^{\infty} \NN(t)e^{-st}\dd t=\NN-+(s),$$
and similarly, $\NN^\sim_+(s)=\NN_-(s)$.


\end{document}